\documentclass[apj,twocolumn,twocolappendix,numberedappendix]{openjournal}
\usepackage{newtxtext,newtxmath,enumitem,multirow}
\usepackage[utf8]{inputenc}
\usepackage[T1]{fontenc}
\usepackage[breaklinks,colorlinks,allcolors=blue,citecolor=blue,urlcolor=blue]{hyperref}
\usepackage{orcidlink}
\pdfoutput=1

\newcommand{\ef}{$\epsilon_\mathrm{ff}$}
\newcommand{\itf}{$1/t_\mathrm{ff}$}
\newcommand{\etf}{$\epsilon_\mathrm{ff}/t_\mathrm{ff}$}
\newcommand{\tf}{$t_\mathrm{ff}$}
\newcommand{\tef}{$t_\mathrm{ff}/\epsilon_\mathrm{ff}$}

\shorttitle{On the universality of star formation efficiency}
\shortauthors{Polzin et al.}
\begin{document}

\title[On the universality of star formation efficiency]{On the universality of star formation efficiency in galaxies\vspace{-1.5cm}}

\author{Ava Polzin\,\orcidlink{0000-0002-5283-933X}$^{1,\star}$}
\author{Andrey V. Kravtsov\,\orcidlink{0000-0003-4307-634X}$^{1,2,3}$}
\author{Vadim A. Semenov\,\orcidlink{0000-0002-6648-7136}$^{4}$}
\author{Nickolay Y. Gnedin\,\orcidlink{0000-0001-5925-4580}$^{1,2,5}$\vspace{2mm}}
\affiliation{$^{1}$Department of Astronomy  \& Astrophysics, The University of Chicago, Chicago, IL 60637 USA}
\affiliation{$^{2}$Kavli Institute for Cosmological Physics, The University of Chicago, Chicago, IL 60637 USA}
\affiliation{$^{3}$Enrico Fermi Institute, The University of Chicago, Chicago, IL 60637 USA}
\affiliation{$^{4}$Center for Astrophysics, Harvard \& Smithsonian, 60 Garden St, Cambridge, MA 02138, USA}
\affiliation{$^{5}$Fermi National Accelerator Laboratory; Batavia, IL 60510, USA}
\thanks{$^\star$\href{mailto:apolzin@uchicago.edu}{apolzin@uchicago.edu}}

\begin{abstract}
We analyze high-resolution hydrodynamics simulations of an isolated disk dwarf galaxy with an explicit model for unresolved turbulence and turbulence-based star formation prescription. We examine the characteristic values of the star formation efficiency per free-fall time, $\epsilon_{\rm ff}$, and its variations with local environment properties, such as metallicity, UV flux, and surface density. We show that the star formation efficiency per free-fall time in $\approx 10$ pc star-forming regions of the simulated disks has values in the range $\epsilon_{\rm ff}\approx 0.01-0.1$, similar to observational estimates, with no trend with metallicity and only a weak trend with the UV flux. Likewise, $\epsilon_{\rm ff}$ estimated using projected patches of $500$ pc size does not vary with metallicity and shows only a weak trend with average UV flux and gas surface density. The characteristic values of $\epsilon_{\rm ff}\approx 0.01-0.1$ arise naturally in the simulations via the combined effect of dynamical gas compression and ensuing stellar feedback that injects thermal and turbulent energy. The compression and feedback regulate the virial parameter, $\alpha_{\rm vir}$, in star-forming regions, limiting it to $\alpha_{\rm vir}\approx 3-10$. Turbulence plays an important role in the universality of $\epsilon_{\rm ff}$ because turbulent energy and its dissipation are not sensitive to metallicity and UV flux that affect thermal energy. Our results indicate that the universality of observational estimates of $\epsilon_{\rm ff}$ can be plausibly explained by the turbulence-driven and feedback-regulated properties of star-forming regions. 
\end{abstract}


%=====================
\section{Introduction}
\label{sec:intro}
%=====================

It has been known for about five decades that the efficiency with which gas is converted into stars in star-forming regions (the star formation efficiency) is small \citep[e.g.,][]{Zuckerman.Evans.1974}. Over the past two decades, observations showed that star formation efficiency per free-fall time is only a few percent for a wide range of environments and scales \citep[e.g.,][see Section 3.2 of \citealt{Krumholz.etal.2019} for a review]{Krumholz.Tan.2007,Krumholz.etal.2012,Garcia.etal.2012,Evans.etal.2014,Utomo.etal.2018,Pokhrel.etal.2021,Hu.etal.2022,Sun.etal.2023,Mattern.etal.2024}.  

Theoretical models aiming to explain the low efficiency on the scales of star-forming regions \cite[see][for reviews]{McKee.Ostriker.2007,Krumholz.2014} include magnetic pressure support against collapse of gas \citep[e.g.,][]{Mouschovias.1976a,Mouschovias.1976b,Price.Bate.2009,Krumholz.Federrath.2019} and supersonic turbulence \citep[e.g.,][]{Elmegreen.2002,Krumholz.McKee.2005}, possibly mediated by magnetic fields \citep[e.g.,][]{Padoan.etal.2012,Federrath.Klessen.2013,Federrath.2015,Girma.Teyssier.2024}. 

On kiloparsec and larger scales galaxies also form stars inefficiently, which is partly related to the local efficiency per free-fall time \citep[e.g.,][]{Krumholz.etal.2012,Semenov.etal.2018}. 
The connection between efficiency in star-forming regions and star formation rate measured on large scales arises because the local inefficiency of star-forming regions and their short lifetimes result in only a small fraction of gas being converted into stars in each star-forming region. Therefore, a given gas parcel has to go through a large number of star formation-dispersal cycles, each taking $\approx 50-100$ Myr as ISM atoms make their way into a star-forming region \citep{Semenov.etal.2017}. 

In this paper, we examine the dependence of the star formation efficiency per free-fall time, {\ef}, on local properties (metallicity, UV radiation field, and density of gas and stars) in a suite of high-resolution simulations that recover observed star formation properties of nearby dwarf galaxies \citep{Kruijssen.etal.2019,Semenov.etal.2021}. We show that simulations naturally produce values of {\ef} in the range $\approx 0.01-0.1$ close to the values estimated for observed star-forming clouds in regions of very different metallicity, UV flux, and density. We show that turbulence and stellar feedback play a key role in producing universal {\ef} values as they regulate the virial parameter of star-forming regions to a fairly narrow range. This is consistent with a picture of self-regulation of star formation via rapid dispersal of star-forming gas in a turbulent ISM.

The paper is laid out as follows: in Section \ref{sec:sim}, we describe the simulation used in this analysis; in Section \ref{sec:results}, we examine the relationship between star formation parameters and local galaxy properties; and in Appendix \ref{sec:turb}, we investigate the role of turbulence in regulating star formation. We discuss and summarize our results in sections \ref{sec:dis} and \ref{sec:conclusions}, respectively.

\begin{figure*}
    \epsscale{1.2}
    \plotone{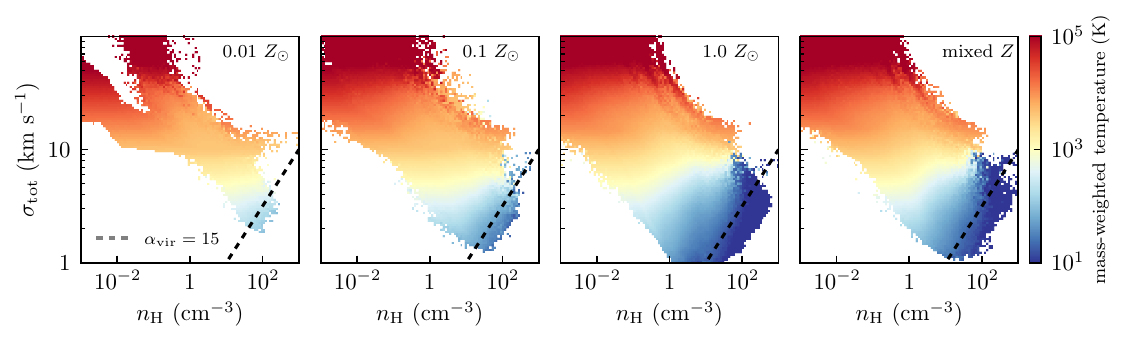}
    \caption{The gas mass-weighted $\sigma_\mathrm{tot}-n$ phase space for four of our simulation runs. The color indicates gas temperature (see side color bar). The figure shows gas with the highest star formation rate (roughly $\alpha_{\rm vir}\lesssim 15$) has temperatures (10-30 K) and densities ($10-1000\,\rm cm^{-3}$), which is similar to the typical temperatures and densities of observed star-forming giant molecular clouds when averaged on similar scales. As discussed in \citet{Polzin.etal.2024}, the low metallicity runs have a lower star formation rate, not because \ef~is intrinsically lower in low $Z$ gas, but because the \textit{amount} of cold, dense gas decreases. 
    \label{fig:stotn}}
\end{figure*}

\begin{figure}
    \epsscale{1.2}
    \plotone{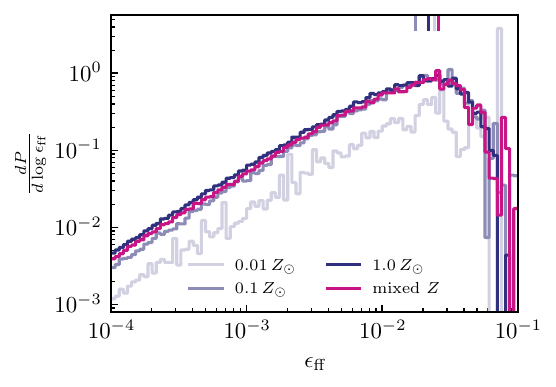}
    \caption{The probability density function of star formation-weighted star formation efficiency per free-fall time in four representative simulation runs. Median values are shown as vertical lines near the top of the figure and demonstrate that there is a $\sim$ universal characteristic \ef~of a few percent in the simulations. The low \ef~tail in $dP/d\log{\epsilon_\mathrm{ff}}$ persists with the same slope even below the range shown here.
    \label{fig:pdf}}
\end{figure}

\begin{figure*}
    \centering
    \includegraphics{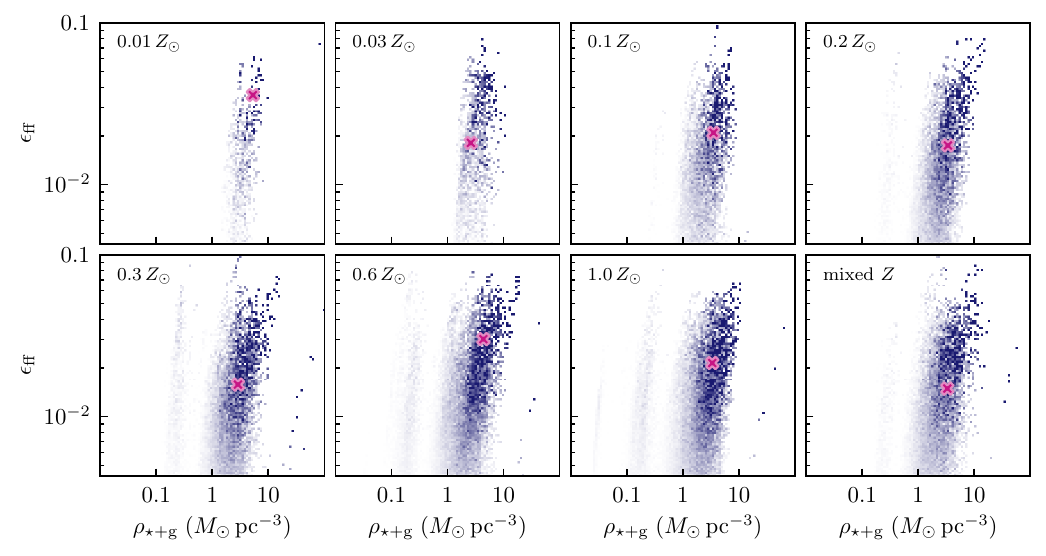}
    \caption{The star formation rate-weighted  star formation efficiency per free-fall time, \ef, as a function of density of gas and stars in individual simulation cells for each of individual runs. The shading indicates the density of the simulation cells with given values of efficiency and density, linearly scaled between the 5th and 95th percentiles of the distribution for each run. Ridges at different densities are associated with different refinement levels in the simulation. Magenta crosses are placed at the star formation rate-weighted median \ef~and $\rho_{\star+\mathrm{g}}$ of the star-forming grid cells.
    \label{fig:voldens}}
\end{figure*}

\begin{figure}
\epsscale{1.2}
\plotone{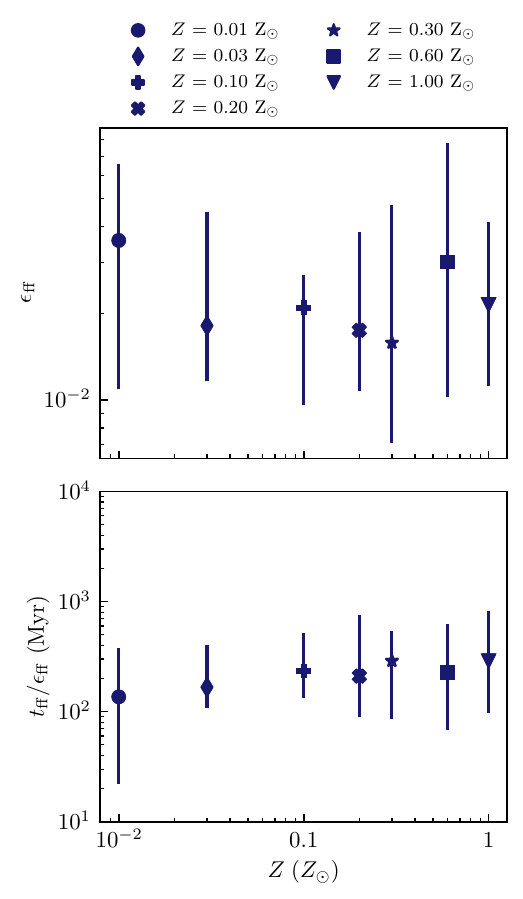}
\caption{The star formation rate-weighted mean $\epsilon_\mathrm{ff}$ (top) and depletion time (bottom) as a function of gas metallicity in the simulation grid cells (i.e. $\sim 10$ pc scale). Error bars correspond to the 16th and 84th percentiles of the distribution.
\label{fig:volZ}}
\end{figure}

\begin{figure}
\epsscale{1.2}
\plotone{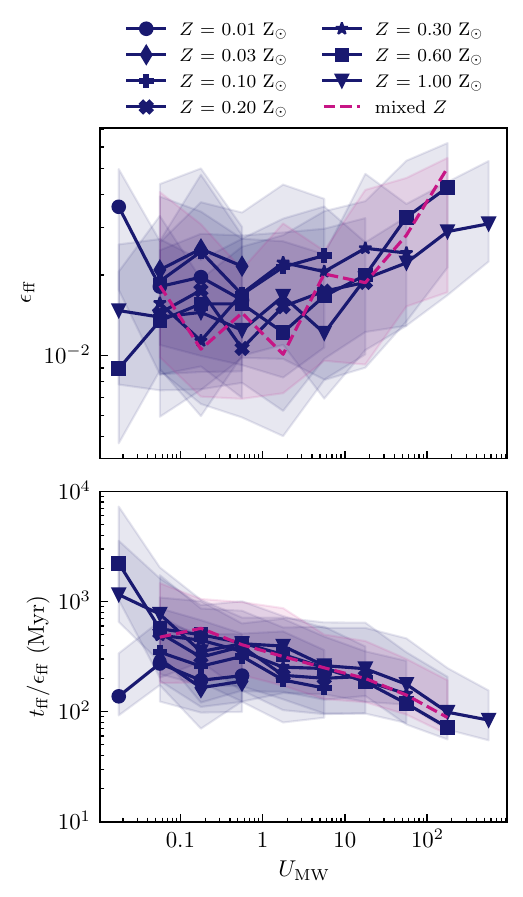}
\caption{The same as Figure \ref{fig:volZ} for the dependence of each run of the simulation on the ionizing radiation field. Shaded regions correspond to the 16th and 84th percentiles of the distribution.
\label{fig:volU}}
\end{figure}

\section{Simulations}
\label{sec:sim}

We use a suite of simulations of an isolated disk galaxy adopting initial conditions similar to the properties of the NGC 300 galaxy \citep{Semenov.etal.2021}. These simulations are run using the ART N-body$+$hydrodynamics code \citep{Kravtsov.1999, Kravtsov.etal.2002, Rudd.etal.2008, Gnedin.Kravtsov.2011} and include self-consistent modeling of radiative transfer (RT; \citealt{Gnedin.2014}), non-equilibrium chemistry of hydrogen, helium, and molecular hydrogen, as well as realistic prescriptions for star formation and feedback. The latter is sufficiently realistic to replicate detailed properties of star forming regions in NGC 300 \citep{Semenov.etal.2021}, and the resolution ($\Delta x = 10$ ~pc in the highest refinement level grid cells, where most star formation is occurring) is sufficient to model feedback-regulated density fluctuations in the simulated interstellar medium.

In these simulations, the star particles form stochastically using the rate $\dot{\rho}_\star=\epsilon_{\rm ff}\rho_{\rm g}/t_{\rm ff}$, where $\rho_{\rm g}$ is gas density, $t_{\rm ff}$ is the free-fall time, and $\epsilon_{\rm ff}$ is the star formation efficiency per free-fall time. The latter is assumed to be an exponential function of the local virial parameter $\alpha_{\rm vir}$ \citep{Padoan.etal.2012}:
\begin{equation}
    \epsilon_\mathrm{ff} = \exp(-\sqrt{\alpha_\mathrm{vir}/0.53})
    \label{eq:eff}
\end{equation}
Here the virial parameter, $\alpha_\mathrm{vir}$, is defined following \citet{Bertoldi.McKee.1992} such that 
\begin{equation}
    \alpha_\mathrm{vir} = \frac{9.35\,(\sigma_\mathrm{tot}/10 \mathrm{\, km\,s^{-1}})^2}{(n/100 \mathrm{\, cm^{-3}})\times (\Delta x/40\, \mathrm{pc})^2}
    \label{eq:avir}
\end{equation}
where the total sub-grid velocity dispersion, $\sigma_\mathrm{tot} = \sqrt{\sigma_\mathrm{t}^2 + c_\mathrm{s}^2}$, includes contributions of both turbulence and the thermal energy of the gas. Note that the virial parameter depends on both the gas temperature through the sound speed and on the turbulent velocity dispersion. This, for example, prevents star formation in the warm gas ($T\gtrsim 10^3$ K). Note also that there is no explicit dependence of $\epsilon_{\rm ff}$ on the Mach number because no significant dependence on the Mach number was found in simulations by \citet[][]{Padoan.etal.2012} and more recent simulations by \citet[][]{Brucy.etal.2024}. Figure \ref{fig:stotn} shows the $\sigma_\mathrm{tot}-n$ phase space in our simulations. The dashed line shows the constant value of $\alpha_\mathrm{vir} = 15$, below which most of the star formation occurs. The figure shows that with the prescription we use in our simulations stars form in cold gas ($T\sim 10-30$ K) with densities of $n\sim 10-10^3\rm\, cm^{-3}$ similar to the densities of observed star-forming molecular clouds averaged on similar scales (tens of parsecs). It also shows that the star-forming gas is different than the gas selected above a constant density and or below a given temperature threshold.

The local value of $\epsilon_{\rm ff}$ in simulation cells evolves with the evolution of the interstellar medium, based on its density and the level of turbulence on small scales. The latter is modeled using an explicit subgrid turbulence model \citep{Semenov.etal.2016}. 

To isolate the dependence of the results on gas metallicity, we ran a series of simulations in which metallicity in all cells is fixed throughout the simulation. Specifically, we ran simulations in which metallicity of all cells was fixed to $0.01Z_\odot$, $0.03Z_\odot$, $0.1Z_\odot$, $0.2Z_\odot$, $0.3Z_\odot$, $0.6Z_\odot$, and $1Z_\odot$ at all times during the run \citep{Polzin.etal.2024}. This is in addition to the original run initialized with the metallicity profile matching the observed radial metallicity profile of NGC 300 and with on-the-fly enrichment turned on \citep{Semenov.etal.2021}. For comparison, we also re-ran these simulations without the subgrid turbulence model, so that the contribution of turbulence to $\alpha_{\rm vir}$ is ignored.

%============================
\section{Results}
\label{sec:results}
%=============================

%---------------------------------------------------------------------------------------------------
\subsection{Star formation efficiency as a function of metallicity on $\sim 10$ pc scale}
\label{sec:props}
%---------------------------------------------------------------------------------------------------

The main result of our analyses is that the values of star formation efficiency per free-fall time in the star-forming regions of our simulations are confined to the range $\sim 0.01-0.1$ with no or little dependence on the local metallicity and UV flux. Figure \ref{fig:pdf} shows the probability density distribution of {\ef} values in the simulations of different gas phase metallicity. The distribution of {\ef} arises from the distribution of $\alpha_{\rm vir}$ values in the simulations. The figure shows that the peak of the PDF and its medians occur in similar ranges at different metallicities. 

Although the distribution extends to very low {\ef} values, the contribution of low-{\ef} regions to the total star formation rate is small. Moreover, results of star-forming cloud simulations of \citet{Padoan.etal.2012} used to set local {\ef} in our simulations were calibrated only at $\alpha_{\rm vir}\lesssim 15$. 
Thus, in the rest of our analysis we focus on the \textit{star-forming} grid cells contributing the bulk of star formation in regions of $\alpha_\mathrm{vir} \leq 15$ or $\epsilon_{\rm ff}\gtrsim 4\times 10^{-3}$.

Figure \ref{fig:voldens} shows the distribution of star formation efficiency per free-fall time as a function of the baryon density (defined here as the density of stars and gas) in our simulation in grid cells (with sizes of $\approx 10-30$ pc). These 2D histograms are weighted by the local SFR in each cell to show the parameter space that contributes most to the total star formation rate. 

The figure shows that the star-formation weighted efficiency per free-fall time has typical values of $\epsilon_\mathrm{ff}\sim 0.01-0.1$, which is similar to the range estimated in observations \citep[e.g.,][]{Krumholz.Tan.2007,Garcia.etal.2012,Evans.etal.2014,Agertz.Kravtsov.2015}, and that the range of efficiencies does not depend on the density. 

We note these characteristic efficiency values are not a result of tuning but arise naturally in simulations (see Section~\ref{sec:dis} for discussion). 
Furthermore, the top panels in Figures \ref{fig:volZ} and Figure \ref{fig:volU} show that the characteristic $\epsilon_\mathrm{ff}$ range does not depend on gas metallicity and depends only very weakly on free-space\footnote{Free-space UV flux is the flux at a given location not attenuated by {\it local} extinction.
 \citet{Krumholz.etal.2009},
for example, use free-space flux to mean the flux incident on molecular clouds. In our simulations, the radiative transfer calculations do not include absorption by H$_2$ lines and thus do not model the radiation field self-consistently inside molecular-rich regions and have to rely on the subgrid model. In this case, the free-space flux is the flux returned by the radiative transfer solver and has the physical meaning of the incident field on the molecular gas.} UV flux, $U_\mathrm{MW}$, normalized by the Milky Way value at 1000\,\AA~\citep{Draine.1978, Mathis.etal.1983}.

Remarkably, while the amount of star-forming gas decreases with decreasing metallicity due to increasingly inefficient cooling \citep[see][for a detailed discussion]{Polzin.etal.2024}, the range of $\epsilon_{\rm ff}$ remains the same for all simulated metallicities. In what follows we show that $\epsilon_{\rm ff}$ also has no, or very weak, dependence on the averaging scale and other properties of the local galactic environment. 

Figures \ref{fig:volZ} and Figure \ref{fig:volU} also show the averages and characteristic range of the local depletion time of the gas in simulations cells, $t_{\rm ff}/\epsilon_{\rm ff}$ as a function gas metallicity and UV flux. Typical depletion time in star-forming cells in the simulation ($\approx 70-500$ Myr) is similar to that measured in the individual star-forming regions in the Milky Way \citep[e.g.,][]{Heiderman.etal.2010,Lada.etal.2010,Lada.etal.2012,Evans.etal.2014}. The figures show that characteristic depletion time does not depend on gas metallicity and exhibits only a weak trend with the UV flux. 

We note that the width of the star-formation weighted rate distribution shown in Figures~\ref{fig:volZ} and Figure \ref{fig:volU} is comparable to the $\approx 0.3$ dex scatter of $\epsilon_{\rm ff}$ estimated by \citet{Hu.etal.2022} for observed star-forming clouds. However, previous studies indicated that the evolution of star-forming cloud properties can contribute significantly to the apparent scatter of $\epsilon_{\rm ff}$ estimated in observations. For example, \citet{Grisdale.etal.2019} show that dramatic changes in the properties of star-forming clouds during their evolution contribute most of the scatter in $\epsilon_{\rm ff}$ values estimated for star-forming molecular clouds the way it is done in observations. Notably, they show that the distribution of $\epsilon_{\rm ff}$ estimated in observations by \citet{Lee.etal.2016} can be matched in a simulation in which intrinsic $\epsilon_{\rm ff}$ in star-forming molecular gas is fixed to a constant value.

\begin{figure*}
\epsscale{1.2}
\plotone{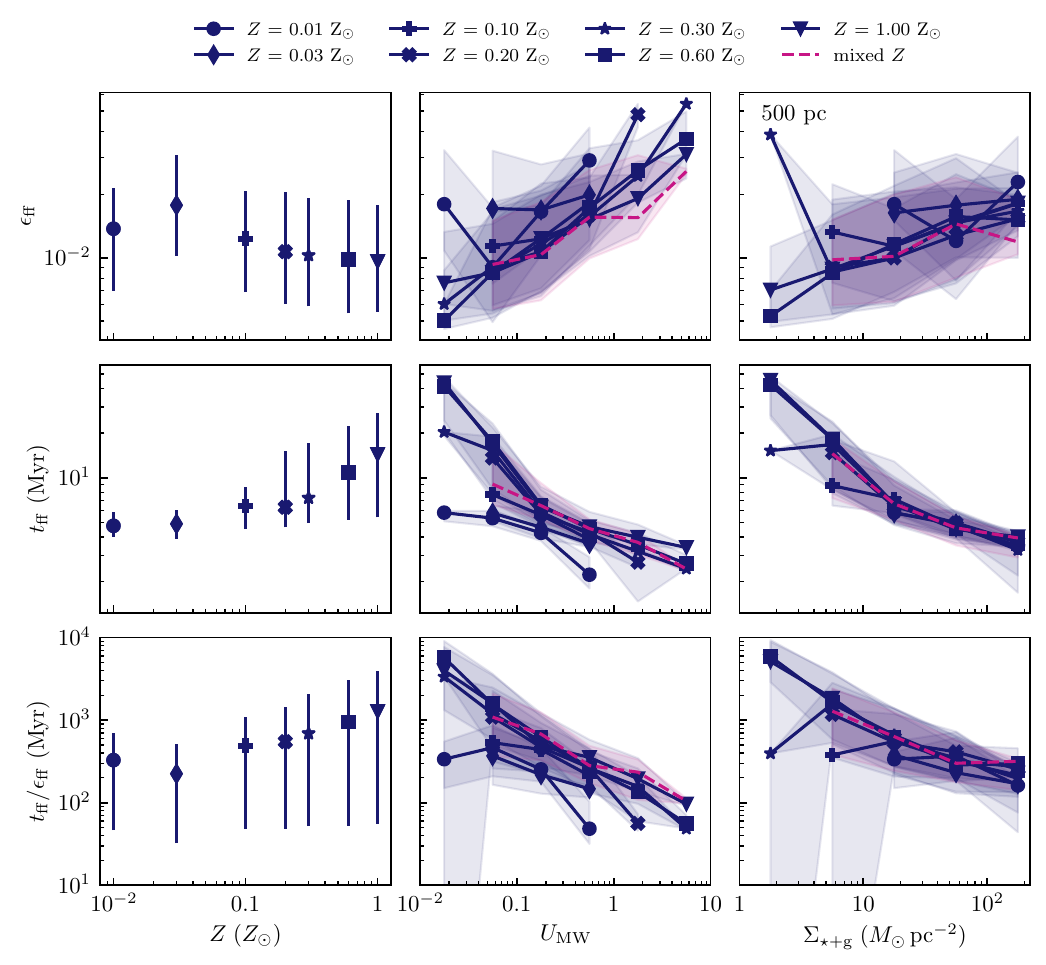}
\caption{\emph{Top:} The star formation rate-weighted median $\epsilon_\mathrm{ff}$ as a function of gas metallicity (left), UV field strength (middle), and surface density of stars and gas (right) smoothed on 500 pc scales for each of our single $Z$ simulation runs and the mixed $Z$ run. \emph{Middle:} The same as the top but for $t_\mathrm{ff}$. \emph{Bottom:} The same as the top but for $t_\mathrm{ff}/\epsilon_\mathrm{ff}$. The error bars and shaded regions correspond to the 16th and 84th percentiles of the distributions. 
% \ak{Error bars in the left panels? What about shaded bands in the center and right column panels?} 
It is notable that \ef, \tf, and \tef~are not dependent on the gas phase metallicity and exhibit only weak trends with UV flux and gas surface density.
\label{fig:summary}}
\end{figure*}

%---------------------------------------------------------------------------------------------------
\subsection{Star formation efficiency as a function of local environment properties on 500 pc scale}
\label{sec:props_proj}
%---------------------------------------------------------------------------------------------------

We examine the star formation efficiency per free-fall time (\ef), the free-fall time (\tf), and the depletion time (\tef), averaged in patches of 500 pc size, as a function of metallicity, UV flux, and baryon (stars and gas) surface density. Namely, we compute the weighted medians\footnote{We define weighted quantiles by the value corresponding to the percentile of the cumulative distribution of weights ordered by the product of the values and the weights.} of these quantities using the normalized local star formation rate in each cell as a weight. 
We also only use the cells with $\alpha_\mathrm{vir} \leq 15$, which contribute the bulk of the total star formation in each simulated galaxy. Note that we actually compute the SFR-weighted averages of \itf~and \etf~because the star formation rate is proportional to these quantities. However, we plot their inverses $t_\mathrm{ff}$ and $t_\mathrm{ff}/\epsilon_\mathrm{ff}$ since the free-fall time and the depletion time values are more easily interpretable.

The left column of Figure \ref{fig:summary} shows the SFR-weighted median $\epsilon_{\rm ff}$, $t_{\rm ff}$, and $t_{\rm ff}/\epsilon_{\rm ff}$ for all star-forming grid cells in 500 pc patches in the runs of different metallicity. 
Similar to the $\epsilon_{\rm ff}$ on $\approx 10$ pc scale, the figure shows that the median efficiency per free-fall time on $500$ pc scale does not exhibit any trend with gas metallicity and ranges in $\epsilon_{\rm ff}\approx 0.007-0.04$ -- the values broadly consistent with efficiencies inferred from observations on similar scales \citep[e.g.,][and references therein]{McKee.Ostriker.2007,Schinnerer.Leroy.2024}. 
The free-fall time and depletion time $t_{\rm ff}/\epsilon_{\rm ff}$ also do not exhibit strong trends with metallicity. 

The middle column of Figure \ref{fig:summary} shows the dependence of \ef, \tf, and {\tef\/} on the UV flux. 
The median {\ef} exhibits a trend with $U_{\rm MW}$, but the trend is very weak with {\ef} increasing by a factor of $\sim$three as UV flux increases by three orders of magnitude. This correlation does not reflect causation, but the fact that regions with higher {\ef} have higher star formation rate and thus higher FUV flux.

Also, over the entire range, the median values of the efficiency are in the range $\approx 0.005-0.05$. Note that this range of values is a bit different than the range of \ef\ in the $\epsilon_{\rm ff}-Z$ panel quoted above because averaging in bins of $U_{\rm MW}$ is over somewhat different gas elements than averaging in bins of $Z$.

There is also a weak anti-correlation of \tf~and $U_\mathrm{MW}$, which can arise because \tf~decreases with increasing density, while $U_\mathrm{MW}$ increases with increasing gas density due to the increase of star formation rate with density.
Finally, there is a weak trend of increasing {\tf} and {\tef} with increasing metallicity at a given $U_\mathrm{MW}$. 

\begin{figure*}
\epsscale{1.2}
\plotone{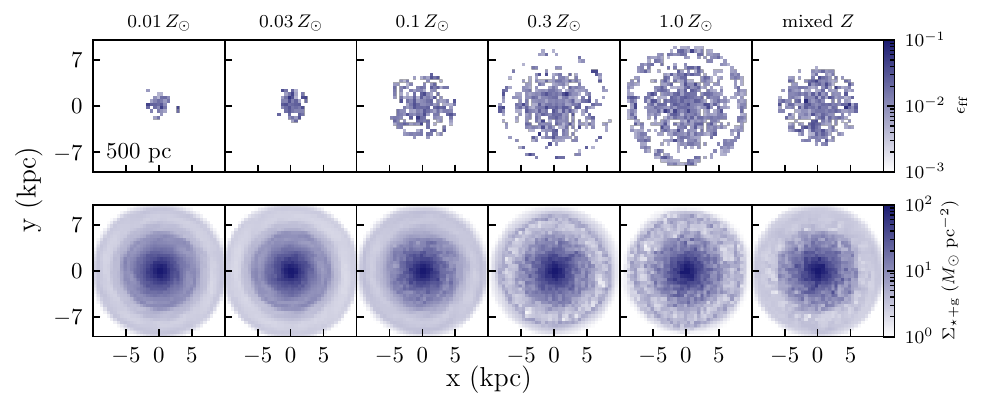}
\caption{Projected face-on maps showing the mean \ef~along the line of sight and the baryon surface density, $\Sigma_{\star\mathrm{+g}}$, on 500 pc scales for some of the runs in the simulation suite we use here.
\label{fig:maps}}
\end{figure*}

The right column of panels of Figure \ref{fig:summary} shows the SFR-weighted medians of {\ef}, {\tf}, {\tef} in the projected patches as a function of the baryon surface density of the patches. The density of gas and stars is directly integrated along the line-of-sight, while the statistics of \ef, \tf, and \tef~are computed using grid cells with $\alpha_\mathrm{vir} \leq 15$ and a refinement level with $\Delta x < 500$ pc. 

The efficiency {\ef} shows only a weak trend with surface density. Our results thus indicate that in the simulations with the turbulence-based SFR model with realistic stellar feedback, the natural result is the universality of characteristic {\ef} values of star-forming gas in environments with a wide range of metallicities, UV fluxes, and densities and over a wide range of averaging scales. This is further illustrated 
in Figure \ref{fig:maps}, which shows the face-on projected maps of {\ef} along with the maps of baryon surface density. The figure visually shows that while surface density varies by more than an order of magnitude in the inner regions of the disk, the values of {\ef} stay in a narrow range around $\epsilon_{\rm ff}\approx 0.01$. \footnote{Note that the trend with surface density shown in Figures \ref{fig:summary} and \ref{fig:maps} is different from that derived by \citet[][]{Sun.etal.2023} for observed galaxies. This is likely because efficiency and free-fall time are defined differently in that study. Namely, the free-fall time is defined as the median time of molecular gas averaged on 60 pc scale, while $\Sigma_{\rm sfr}$ in their study is estimated on the 1.5 kpc scale rather than 500 pc in our case.  Moreover, we define both {\ef} and $t_{\rm ff}$  relative to the star-forming gas, which is not the same as molecular gas in our model \citep[see][for detailed discussion]{Semenov.etal.2019}.
If our definition $\epsilon_{\rm ff}\approx \Sigma_{\rm sfr}/(\Sigma_{\rm sf}/t_{\rm ff})$, where $\Sigma_{\rm sf}$ is the surface density of star-forming gas, while \citet{Sun.etal.2023} define $\epsilon_{\rm ff,mol} = \Sigma_{\rm sfr}/(\Sigma_{\rm mol}/t_{\rm ff,mol})$, where $\Sigma_{\rm mol}$ is the surface density of molecular gas. The two definitions are thus related by 
$\epsilon_{\rm ff} = (\Sigma_{\rm mol}/\Sigma_{\rm sf})(t_{\rm ff}/t_{\rm ff,mol}) \epsilon_{\rm ff,mol}$. The relation depends on the trend of $(\Sigma_{\rm mol}/\Sigma_{\rm sf}) (t_{\rm ff}/t_{\rm ff,mol})$ with surface density.}

The median free-fall time of star-forming gas, on the other hand, shows tight, nearly metallicity-independent correlations with the UV flux and baryon surface density. This indicates that star-forming gas is denser in denser regions of the galaxy. This may be because gas in high surface density regions experiences a stronger gravitational pull by the disk compressing it to higher densities. The trend with UV flux may reflect the positive correlation between the UV flux and star formation rate, which, in turn, correlates with gas density. These trends of {\tf} result in similar trends of the depletion time with $U_{\rm MW}$ and $\Sigma_{\rm \star+g}$.

%=======================
\section{Discussion}
\label{sec:dis}
%=======================

As noted above, the characteristic values of \ef in our simulations arise naturally from the specific star-formation prescription based on the simulations of star formation in turbulent molecular clouds \citep[see eq.~\ref{eq:eff} above;][]{Padoan.etal.2012}. The efficiency in this simulation-calibrated prescription is an exponential function of $\sqrt{\alpha_{\rm vir}}$, where $\alpha_{\rm vir}$ is the virial parameter given by eq.~\ref{eq:avir}, which measures the relative importance of turbulent and thermal pressure relative to gravitational force. 

The characteristic values of $\epsilon_{\rm ff}\approx 0.01-0.1$ measured in the simulations thus correspond to the characteristic values of $\alpha_{\rm vir}\approx 3-10$, which are set by the dynamics of the ISM gas and stellar feedback that affects the thermal and turbulent pressure of the gas. 

Large-scale ISM flows lead to compression and expansion of the gas and fluctuations of $\alpha_{\rm vir}$ \citep[e.g.,][]{Semenov.etal.2017}. As a local compression decreases $\alpha_{\rm vir}$ locally, {\ef} grows as $\propto \exp(-\sqrt{\alpha_{\rm vir}})$. This growth is accompanied by the growth of star formation rate and associated stellar feedback, which increases local thermal and, importantly, turbulent pressure. This limits the decrease of $\alpha_{\rm vir}$ and leads to self-regulation of its values to the characteristic range of  $\alpha_{\rm vir}\approx 3-10$. Part of the reason behind the narrow range of $\alpha_{\rm vir}$ is the fact that during compression and expansion due to large-scale flows and feedback turbulence often evolves adiabatically, $\sigma\propto n^{1/3}$, and the virial parameter changes only mildly during such evolution, $\alpha_{\rm vir} \propto \sigma^2/n \propto n^{-1/3}$.
The characteristic values of $\epsilon_{\rm ff}\approx 0.01-0.1$ in our simulations are thus the result of the star formation prescription and the stellar feedback model adopted in our simulations.

This also allows us to interpret the universality of {\ef} in environments of different metallicity, UV flux, and densities demonstrated above. The pressure and $\alpha_{\rm vir}$ in star-forming regions in these simulations are dominated by turbulence. The turbulent energy is driven by gas compression by large-scale flows and the dissipation on the turbulent crossing time scale.
Both processes do not depend on the metallicity, UV flux, or density.  In contrast, the gas cooling and heating, and therefore thermal pressure, do depend on these properties and thus, when $\alpha_{\rm vir}$ is dominated by 
the thermal pressure, the efficiency per free-fall time varies more strongly with the properties of the environment.
This is illustrated explicitly in Appendix \ref{sec:turb}, where we show {\ef} as a function of metallicity, UV flux, and surface density in simulations in which subgrid turbulence is not modeled and its contribution to $\alpha_{\rm vir}$ is ignored. The turbulence thus plays an important role in making {\ef} value range universal.  

The values of {\ef} and their universality exhibited in our simulations are consistent with the typical values of $\approx 0.01-0.1$ estimated for observed gas in diverse environments and averaging scales \citep[e.g.,][]{Garcia.etal.2012,Utomo.etal.2018,Sun.etal.2023}. As noted in the previous section, the definitions of star formation efficiency and scales on which it is measured vary among studies. The comparison here is therefore ballpark by necessity and the trends with scale or surface density for different {\ef} definitions may differ. 

Estimates of \ef~ from the star-forming sub-pc cores to giant molecular clouds on tens of pc scales are $\approx 0.01-0.1$ \citep[e.g.,][]{Krumholz.Tan.2007,Evans.etal.2014,Pokhrel.etal.2021,Hu.etal.2022}. Similar values of {\ef} are estimated on the scales of hundreds of pc in other galaxies with a diverse range of metallicities and surface densities \citep[e.g.,][]{Leroy.etal.2008,Utomo.etal.2018,Teng.etal.2024}. Our results indicate that these characteristic values of {\ef} and their universality can be plausibly explained by the turbulence-driven and feedback-regulated properties of star-forming regions. 

We note, however, that comparing detailed distributions of {\ef} estimated in observations and simulations is not straightforward. In simulations, instantaneous values of {\ef} are considered, while in observations, these values are estimated using instantaneous gas properties and star formation rate indicators that are sensitive to star formation over a certain time scale. The gas density is expected to evolve strongly as the star formation and feedback proceed in a given region \citep{Feldmann.Gnedin.2011}. The large apparent scatter of observational estimates of {\ef} thus largely reflects diverse evolutionary stages of the regions in which these estimates are made \citep[e.g.,][]{Feldmann.Gnedin.2011,Lee.etal.2016,Grisdale.etal.2019}. Nevertheless, the simulations shed light on why the characteristic average values of {\ef} do not vary with environment properties. 

Finally, we note that recently \citet[][]{Semenov.etal.2024} and \citet[][]{Segovia_Otero.etal.2024} presented results of cosmological galaxy formation simulations of Milky Way-sized galaxies with the same prescription for {\ef} given by equations 1 and 2 above. In both studies, the authors found that the average {\ef} of the simulated galaxies was $\sim 1\%$, although rapidly changing conditions during early stages of evolution result in significant efficiency fluctuations around this characteristic value. In particular, \citet[][]{Semenov.etal.2024} showed that fluctuations of {\ef} are the dominant source of burstiness of star formation during the early stages of galaxy evolution.

%================================
\section{Summary and conclusions}
\label{sec:conclusions}
%================================

In this study, we analyze high-resolution simulations of gas dynamics and star formation in an isolated disk dwarf galaxy with an explicit model for unresolved turbulence and turbulence-based star formation prescription calibrated on MHD simulations of turbulent molecular clouds. As demonstrated by \citet{Semenov.etal.2021}, the star formation and feedback model used in the simulations reproduces star formation and gas correlations in observed galaxy NGC 300 in fine detail. We use a suite of simulations at different metallicities to examine the characteristic values of the star formation efficiency per free-fall time, {\ef}, and its variations with local environment properties, such as metallicity, UV flux, and surface density. Our results and conclusions can be summarized as follows. 
\begin{itemize}
    \item[1.] We show that the star formation efficiency per free-fall time in $\approx 10$ pc star-forming regions of the simulated disks has characteristic values in the range $\epsilon_{\rm ff}\approx 0.01-0.1$ (Figures \ref{fig:pdf} and \ref{fig:voldens}). We show that the characteristic values of {\ef} do not vary with metallicity and exhibit only a very weak trend with UV flux  (Figures \ref{fig:volZ} and \ref{fig:volU}).
    \item[2.]  Likewise, we show that {\ef}  estimated on a $500$ pc scale does not vary with metallicity and shows only weak trends with average UV flux and gas surface density estimated at this scale (Figures \ref{fig:summary} and \ref{fig:maps}).
    \item[3.] The characteristic values of $\epsilon_{\rm ff}\approx 0.01-0.1$ arise naturally in the simulations as a result of the gas dynamics leading to gas compression and ensuing stellar feedback that injects thermal and turbulent energy into the gas. The compression and feedback regulate the virial parameter, $\alpha_{\rm vir}$, in star-forming regions, limiting it to the range $\alpha_{\rm vir}\approx 3-10$ (see discussion in Section~\ref{sec:dis}). Turbulence plays an important role in the universality of $\epsilon_{\rm ff}$ because turbulent energy driven by gas dynamics and feedback and its dissipation are not sensitive to the standard metallicity and UV flux cooling and heating processes that affect thermal energy. 
    \item[4.] We show explicitly that $\epsilon_{\rm ff}$ values do depend on the metallicity and UV flux in simulations where turbulent pressure contribution to the virial parameter is ignored (Figure \ref{fig:noturb}).
\end{itemize}

The efficiency per free-fall time and its universality are similar to the values of observed star-forming regions in different environments and galaxies. This indicates that the universality of observational estimates of {\ef} can be plausibly explained by the turbulence-driven and feedback-regulated properties of star-forming regions. In the future, it will be important to forward model observational measurements using simulations taking into account evolutionary effects on the estimates of {\ef}. 

\begin{acknowledgements}
The authors thank the referees for comments that improved the manuscript.
A.K. was supported by the National Science Foundation grant AST-1911111 and NASA ATP grant 80NSSC20K0512.
V.S. is grateful for the support provided by Harvard University through the Institute for Theory and Computation Fellowship.
The simulations used in this work were carried out on the Midway cluster maintained by the University of Chicago Research Computing Center. Analyses presented in this paper were greatly aided by the following free software packages: {\tt NumPy} \citep{harris2020array}, {\tt SciPy} \citep{scipy}, {\tt Matplotlib} \citep{Hunter:2007}, {\tt AstroPy} \citep{astropy:2013, astropy:2018, astropy:2022} and {\tt yt} \citep{Turk.etal.2011}. We have also used the Astrophysics Data Service (\href{http://adsabs.harvard.edu/abstract_service.html}{\tt ADS}) and \href{https://arxiv.org}{\tt arXiv} preprint repository extensively during this project and the writing of the paper.
\end{acknowledgements}

\appendix
\section{Turbulence-free star formation}
\label{sec:turb}

To better understand the mechanism driving the uniformity of the star formation efficiency with varying ISM properties (metallicity, UV field strength, and surface density of gas and stars), we re-run five of the single metallicity simulations ($Z = 0.01, 0.03, 0.1, 0.3,$ and $1\, Z_\odot$) from the same initial conditions as the fiducial simulations introduced in Section \ref{sec:sim}, turning off subgrid turbulence and any tracking of turbulent energy. The non-turbulent virial parameter, $\alpha_\mathrm{vir, nt}$, is then defined taking $\sigma_\mathrm{tot} = c_\mathrm{s}$, rather than $\sigma_\mathrm{tot} = \sqrt{\sigma_\mathrm{t}^2 + c_\mathrm{s}^2}$ as in the default \citet{Padoan.etal.2012} model (see Eq. \ref{eq:avir}), effectively isolating the behavior and effects of thermal motions in star-forming gas. After the runs each proceed for $\sim 100$ Myr (approximately a crossing time), \ef~is then recomputed using $\alpha_\mathrm{vir,nt}$ for the non-turbulent case. We denote this non-turbulent star formation efficiency per free-fall time as $\epsilon_\mathrm{ff,nt}$. Similarly, the turbulence-free star formation rate is defined as $\dot{M_\star} = \epsilon_\mathrm{ff,nt} M_\mathrm{gas}/{t_\mathrm{ff}}$.

\begin{figure*}
    \centering
    \includegraphics{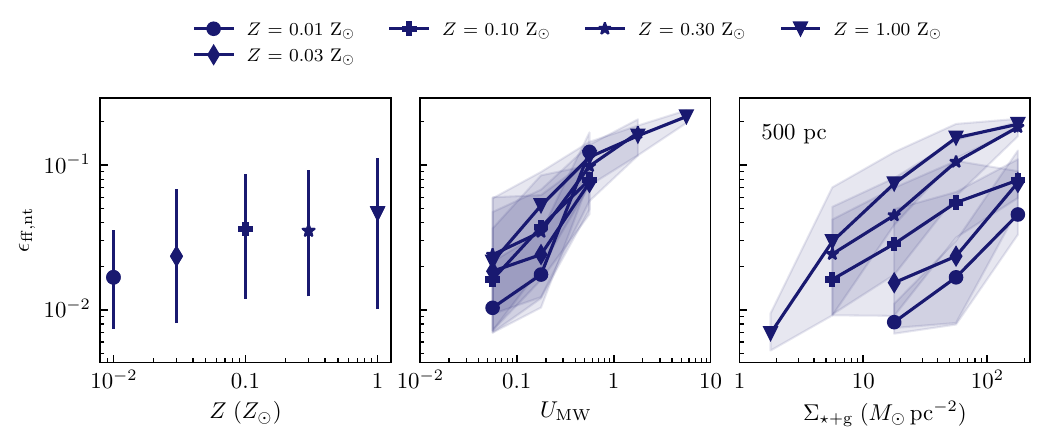}
    \caption{The same as the top row of Figure \ref{fig:summary}, but with turbulence turned off. The now-strong dependence of the star formation efficiency on metallicity and baryon surface density is apparent. (Note that the range of \ef~shown here is larger than in Figure \ref{fig:summary}.) 
    }
    \label{fig:noturb}
\end{figure*}

All other analysis is done in the same way as in Section \ref{sec:props}, and we do not change any of the physics associated with star formation or feedback within the simulation. For direct comparison between Figure \ref{fig:summary} and \ref{fig:noturb}, we make the same cut of $\alpha_\mathrm{vir} \leq 15$, with cells weighted by their star formation rates.

The dependence of $\epsilon_\mathrm{ff,nt}$ on gas phase metallicity is shown on the left in Figure \ref{fig:noturb}. In the turbulence-free case, \ef~dependence on metallicity is noticeably stronger than in the fiducial simulations. This is likely due to the importance of thermal motions in defining $\alpha_\mathrm{vir,nt}$. Even selecting only for star-forming regions ($\alpha_\mathrm{vir} \leq 15$), low metallicity gas will be warmer than higher $Z$ gas given the less efficient cooling in low metallicity gas. In this case, the thermal motions defined by $c_\mathrm{s}$ will be greater in low $Z$ gas, increasing $\alpha_\mathrm{vir,nt}$ and driving $\epsilon_\mathrm{ff,nt}$ down compared with $\epsilon_\mathrm{ff}$. This leads to \ef~varying increasing steadily with increasing gas phase metallicity.

Likewise, Figure \ref{fig:noturb} shows that in the simulations without subgrid turbulence {\ef} depends significantly on the local UV flux and gas surface density (the middle and right panels). The middle panel of Figure \ref{fig:noturb}. The figure shows that $\epsilon_\mathrm{ff,nt}$ increases with increasing $U_\mathrm{MW}$ and surface density of gas and stars (averaged on 500 pc) due to the positive correlation between $U_\mathrm{MW}$ and density and star formation rate. The significant dependence of the efficiency on metallicity at specific values of $U_{\rm MW}$ and surface density are also apparent in these panels. 

Note that stellar feedback still operates in the runs without turbulence. Thus, as gas cools and reaches higher densities star formation rate increases and so does associated stellar feedback. The feedback prevents runaway collapse and disperses star-forming regions.

\bibliography{references}{}
\bibliographystyle{aasjournal}

\end{document}